\documentclass[%
 reprint,
 amsmath,amssymb,
 aps,
floatfix,
]{revtex4-2}
\usepackage{lipsum}
\usepackage{graphicx}% Include figure files
\usepackage{dcolumn}% Align table columns on decimal point
\usepackage{bm}% bold math
\begin{document}

\preprint{APS/123-QED}

\title{On the Automation of High Throughput Modeling of Adsorption In Porous Zeolitic Imidazolate Frameworks}% Force line breaks with \\

\author{Michael Atambo}
 \altaffiliation[Also at ]{Kenya Education Network Trust.}
\affiliation{%
 Technical University of Kenya, Nairobi, Kenya.
}%
\homepage{https://spas.tukenya.ac.ke/index.php/20-materials-modeling-group-department-of-physics-and-space-sciences}
\author{Kiptiemoi Kiprono Korir}% 
\affiliation{
 Moi University, Eldoret, Kenya
}%

\date{\today}% It is always \today, today,

\begin{abstract}
%\lipsum[2-2]
The zeolitic imidazolate frameworks (ZIF) have emerged as a promising candidate for catalysis, carbon-dioxide (CO$_{2}$) capture and storage as well as flue gas separation due to their tunable porosity and chemical stability.  ZIFs consists of transition metals in a tetrahedral coordination with imidazolate linkers, allowing for structural modifications that can enhance CO$_{2}$ adsorption and storage. To systematically design and optimize ZIFs for superior CO$_{2}$ capture performance, theoretical modeling provides an effective approach, though hindered by the sheer size of the search space, with hundreds to thousands of possible sites per ZIF, and thousands of possible ZIF that can be synthesized. In this work, we employ a high-throughput techniques underpinned by first-principle Density Functional Theory (DFT) to develop a systematic automated method for characterizing  adsorption energetics in ZIFs, guiding the rational design and selection of desirable ZIF structures. Using high-throughput computing tools, we perform pore and pore size analysis, and implement automated CO$_{2}$ molecule placement algorithm, accounting for positional symmetry, crystallographic orientation, and collision detection. The results obtained are in good agreement with previous studies, demonstrating the reliability of the approach in accelerating the discovery of next-generation ZIFs for CO$_{2}$ capture and storage.

%\begin{description}
%Secondary publications and information %retrieval purposes.
%\end{description}
\end{abstract}

\maketitle

%\tableofcontents

\section{\label{sec:level1} Zeolitic Imidazolate Frameworks (ZIF) }
\subsubsection{\label{sec:level2C}Organic Frameworks}
MOFs, COFs and ZIFs have attracted a strong interest \cite{zheng2023structural}, and advancements in modelling them\cite{cho2024improving} and have tremendous 
commercial promise, due to their chemical and thermal stability, together with 
their porosity, adsorbtion capacity, selectivity\cite{long2009pervasive,frivsvcic2014environmentally},  ionic conductivity
and glassforming/hybrid glasses\cite{leon2023meltable}, sensing\cite{kukkar2021recent} and catalysis\cite{barona2019computational,rosen2019structure, xiao2014oxidation}. 
ZIFs, have been the target of synthesis and characterization \cite{zheng2023structural}, 
including using rapid high throughoughput synthesis \cite{banerjee2008high}. Due to large scale release
of carbon dioxide from anthropogenic sources, efforts are required to address the
sequestration of the CO2 gases, and limit the downstream consequences of the 
green house effect. 

\subsubsection{\label{sec:level2A}ZIFs}
ZIFs are considered among the best candidates for CO2 capture and selective separation via the means of physisorption\cite{phan2009synthesis, gallouze2016adsorption}, exhibiting high gas affinity, capacity, 
complete reversibility in their uptake isotherms, and importantly comparatively
better selectivity for CO2 gas than COFs/MOFs. From a design point of view, ZIF's offer an additional set of advantages, which include control over the pore aperture, the
framework chemistry, surface area and pore volume\cite{pimentel2014zeolitic,li1999design,furukawa2013chemistry,freund2021current,wang2021progress}. 

Adsorbtion sites within ZIFs are  associated with the Ligand and metal center, and are a 
consequence of the topology, chemical nature of the ligand, the partial pressure of the system
as and other factors \cite{pimentel2014zeolitic,cho2024improving}. By construction, ZIFs consist of transition
metals with four-fold coordination through imidazolate units, creating an extended tetrahedral
framework topology, with complex interlinking that creates cages\cite{phan2009synthesis}. It is therefore
possible that from the diverse array of possible imidazolate building units, a candidate with 
unsurpassed gas adsorption properties can be produced from the large class of possible structures,
and effort that is underway through both small scale and high throughput
synthesis\cite{chen2014zeolitic,leon2023meltable}.
\subsubsection{\label{sec:level2B}Theoretical Modeling of ZIFs}
Theoretical modeling is imperative to explore this space of ZIFs, since it is very large, however,
in addition to this, the chemical environment can be tailored though the choice of metal cations 
and ligands, creating a need for first principles study to characterize the metal-linker-adsorbate 
interactions, and develop the capacity to design selectivity for specific gases. We are therefore
at a point where a high throughput first principles exploration of ZIF-adsorbate is both required 
and possible, even with the extreme computational needs of studying large molecules at this level 
of theory. 

Prior theoretical work on ZIFs has included a mix of earlier classical \cite{yu2017co2,ray2012van,liu2011molecular,chen2012effects}, and recent first principles DFT approaches\cite{paudel2021computational, gallouze2016adsorption,safia2017theoretical,hoang2018reticular}, and developments that address known shortcomings of DFT using DFT+U \cite{cho2024improving},
 which is both more accurate and computationally expensive. 

Therefore, using the tools of \emph{high throughput computing} (HTC) to solve the computational challenge of ZIFs,
we have developed an automated workflow for the first principles characterization of the energetics
of adsorption in ZIFs using DFT, and, being a generic workflow, can be used with DFT adjacent 
techniques such as DFT+U and Hybrids, to answer the question of computational modeling in the 
exascale era, with 1000s of atoms per system with 1000s of possible sites per system.

\section{Methods:  Automated Adsorbed Structure Generation.}
To generate the adsorbed structure realizations, we required systematic and detailed pore geometry,
including the spatial location within the cell, the pore's geometric center and pore size, and all
of this information in a programmatic format suitable for further analysis. Once this task was completed
the pore information was used to create a set of adsorption positions where the molecule would be placed.
The former of this task we performed using the code porE\cite{trepte2021pore}, the later tasks were implemented as part of this
work as described here.

\subsection{Pore analysis}
  PorE\cite{trepte2021pore} provides, among others, a monte-carlo based implementation of pore size distribution analysis, implemented in
 Fortran and available through a python interface. The pore size distribution is achieved via sampling
 technique starting from random positions within the cell :
  $\mathbf{r}_i = \alpha\cdot \mathbf{a} + \beta\cdot \mathbf{b} + \gamma\cdot \mathbf{c}, $
 where $\alpha, \beta$ and $\gamma$ are random numbers, while $ \mathbf{a}, \mathbf{b}$ and $ \mathbf{c}$
are the cell vectors. A random walk throughout the entire cell, respecting periodic boundary conditions
then is used to construct a complete picture of the pore size distribution within the cell. PorE was used for pore volume analysis, yielding the pore center coordinates, the pore volume and radius $R$.

\subsection{Automated Molecule Placement}
Given the geometric pore size distribution analysis data, a followup procedure is required to compute all
possible center of mass (COM) positions to place the adsorbed molecule in a pore, that do not violate several
conditions, including, overlapping with host atoms and overlapping with other positions that require sampling.
In this process, the CO2  is treated as a linear molecule. In addition to the center of mass positions,
multiple orientations are also considered at each COM, these orientations are obtained via a symmetry
and crystal direction analysis.

\subsubsection{Position Sampling}
The pore volume is sampled through set of uniformly distributed $NP$ positions, every $l=L/N$th of the unit cell length $L$, in a sphere with the diameter of the void $2R$, centered at the geometric center of the void,
$$ NP \propto R^3/l^3  ,$$
achieved using  $X_i$, $Y_i$ and $Z_i$ as independent standard normal distributions with $NP$ points, and,
$$ \frac{RU^{1/3}}{\sqrt{ |X|^2 + |Y|^2  + |Z|^2 } } (X,Y,Z), $$
where $U$ is uniformly sampled between $[0,1]$,
where each of $X(x_i)$ are independent distributions for the $x$, $y$ and $z$ coordinates,
$$X(i) \in \frac{\frac{1}{\sqrt{2\pi}} e^-{\frac{x_i-\mu}{2\sigma}} } { ||x_i|| } .$$

However, the interior wall of the pore may not be necessarily uniform, and its shape will therefore not be a perfect sphere.
This has to be accounted for to ensure clefts and pockets in the pore are well sampled, even when beyond the
average radius obtained from the pore size distribution analysis.

  \subsubsection{Reverse Onion partially Occluded inclusion.}
\begin{figure}
\includegraphics[height=50mm,width=.60\columnwidth]{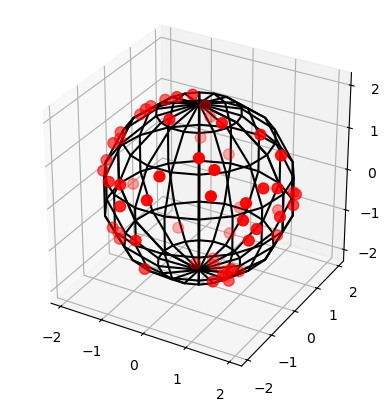}%
\caption{\label{fig:sampling} Uniform sampling on a layer of the surface of a sphere.}
\end{figure}

This is resolved by considering a reverse onion algorithm, that appends a new layer of atoms on the surface of the previously
sampled sphere. The surface of the sphere is sampled through a normal distribution over the sphere of radius $R+ \delta$, holding
the radius at $R+\delta$, where we step outwards adding a new layer to the sphere, $R+\delta_i$, where, $\delta = 1/L $,

$$ \frac{R}{\sqrt{ |X|^2 + |Y|^2  + |Z|^2 } } (X,Y,Z), $$
normalizing the distribution to a magnitude of unity, where each of $X(i), Y(i),Z(i)$ are independent distributions for the $x$, $y$ and $z$ coordinates.
Once the new layer added, all the positions are kept or rejected, ensuring that there is no overlap with the host atom positions
or other sampled positions, this is performed for each added position on the surface of the sphere. This reverse onion procedure is
repeated until all positions on the sphere are rejected for the entire new onion layer. This algorithm ensures we sample pockets and
irregular shapes uniformly without omitting volumes beyond the average radius of the pore. This procedure was implemented in this work. A plot of the sampling on the surface of the sphere is shown in Fig.~\ref{fig:sampling}

\subsubsection{Symmetry and Direction Sampling}
\begin{figure}
\includegraphics[height=57mm,width=.70\columnwidth]{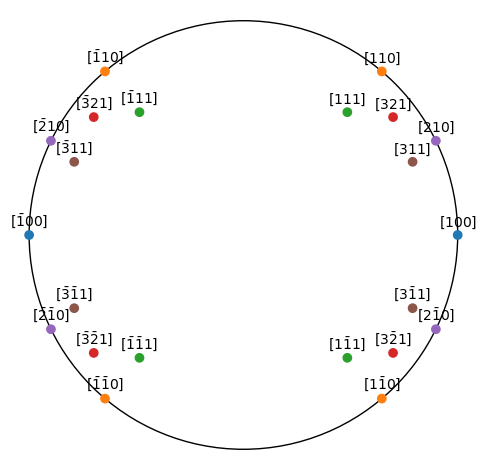}%
\caption{\label{fig:wide-hemisphere} Symmetrically inequivalent directions for families in the $mmm$ point group of the orthorombic system over the upper hemisphere projected on a disk.}
\end{figure}

Given CO$_2$ is treated as a linear molecule withing the ZIF crystal,  we can obtain from the symmetry operations $S$ on the point group symmetry of the lattice, a description of  symmetrically unique crystal directions $[u,v,w]$ on the upper hemisphere. A set of directions was selected after Ray et. al.\cite{ray2012van}. 
This analysis reduces the number of possible orientations of the molecule by  eliminating symmetrically equivalent crystal directions located at a particular molecule COM position that has to be considered. This is available in the Orix package\cite{johnstone2020density,hakon_wiik_anes_2025_14601499}. Fig.~\ref{fig:wide-hemisphere} shows the symmetrically inequivalent directions for ZIF-CO3-1.  

\subsubsection{Collision Detection}
The atomic position overlap detection/collision detection on the structure was implemented in this work using a nearest neighbor approach implemented using a KDTree data structure. A molecule position is rejected if it falls within the covalent radii of the nearest neighbor. 

\section{Methods: High Throughput Workflow.}
Using the high throughput and reproducible science framework AiiDA\cite{huber2020aiida,uhrin2021workflows}, we made use of the \texttt{PwBaseWorkChain} to automate job execution using the AiiDA QE plugin\cite{huber2020aiida,uhrin2021workflows}, and the AiiDA python API for further analysis. The automated error handling, resubmission and data provenance afforded by AiiDA enabled the scalability of this work to large numbers of possible structural realizations for the adsorbed molecule. The uniformity of the API means that the underlying engine, and directly the level of theory targeted can be easily swapped for different levels of accuracy, including classical (Car-Parinello) as well as DFT or other higher order methods supported by AiiDA's quantum engines. 

\section{\label{sec:results} Results and Analysis.}
\subsection{Automated Adsorbed Structure Generation.}
Tab.~\ref{tab:table3} shows the results of the structural sampling within a pore, and the resulting unique possible adsorption site. The introduction of symmetry considerations and minimum nearest neighbor rules reduces the number of realizations significantly, without impacting the efficacy of the sampling within the pore volume. 

\begin{table*}
\caption{\label{tab:table3} Number of structural realizations for different ZIF frameworks and choice of N, where N is used to choose the step $\delta$ as a fraction of the length of the cell $L$ }
\begin{ruledtabular}
\begin{tabular}{lccccc}
 ZIF& Space group & point group & N & Realizations & Reduced Realizations (by Symmetry) \\ \hline
 ZIF$-$CO3-1 & Pba2 & $mmm$ & 24 & 4586 & 1555 \\
             & Pba2 & $mmm$ & 48 & 20476 & 7164 \\
\end{tabular}
\end{ruledtabular}
\end{table*}

\subsection{High Throughput Calculations}
However, that said, it is generally the case that the number of realizations will be considerable, with ZIF-CO3-1\cite{basnayake2015carbonate,zheng2023structural}, selected as the model system,  having one of the lowest pore volumes, therefore also the lowest number of realizations. For larger frameworks with larger pore volumes, this number is magnitudes larger, and as a result, poses a robust high throughput computing challenge, with thousands of single point calculations required to systematically characterize a single pore volume.

The separation of the structure generation and storage step, made it possible to submit the same in a scalable high throughput approach, that queried for the structures, submitted and stored the results in the AiiDA database. 

\subsection{ZIF-CO3-1 Energetics and Adsorption}
\begin{figure}
\includegraphics[height=65mm,width=.9\columnwidth]{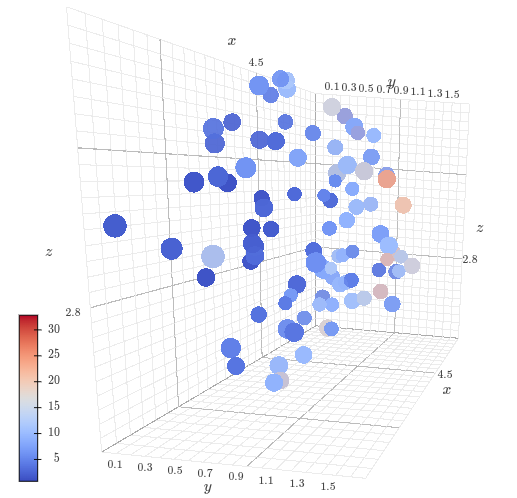}%
\caption{\label{fig:no-context} The adsorption energies at a set of uniform random points within the pore volume of ZIF-CO3-1 showing the gradient between the energetics between the most unfavorable (orange), to the lease unfavorable (blue) across the pore void.}
\end{figure}

To understand the thermodynamic stability of CO$_2$ adsorption in ZIF-CO3-1 it is useful to compute the binding energy, which we define as such: 
$$ E_b = E_{ZIF+CO2} - (E_{ZIF} + E_{CO2}) $$
where $ E_{ZIF+CO2}$ is the total energy posterior to the adsorption,  while $E_{ZIF}$ and $E_{CO2}$ correspond to the individual total energies prior to the adsorption, and for $E_{CO2}$, the molecule is placed in a cell of the same size as the Framework's unit cell. Fig.~\ref{fig:no-context} shows the resulting adsorption energies, in eV. 

From an examination of these energies, it is apparent that there is a general favorability of the adsorption energies as the CO$_2$ moves further from the cusp of the pore. The $>$ 0 adsorption energies, however, in line with Basnayake \cite{basnayake2015carbonate}, indicate the primary mechanism is reactive binding rather than conventional adsorption. 

%https://sci-hub.se/https://doi.org/10.1063/5.0054874

\section{Conclusion}
The use of systematic sampling, orientation analysis and symmetry analysis provided a means to sample all the desired positions within the pore volume of ZIF-CO3-1, reducing the computational burden by an order of magnitude.
From an analysis of the results, we identified the most energetically favorable positions for adsorption to occur, and found conventional adsorption to be less favorable than reactive binding. 
Furthermore, the use of the AiiDA framework ensures scalability by automating error handling, job submission and resubmission as well as data analysis, provenance recording and reproducibility.

%The equation that follows is set in a wide format, i.e., it spans the full page. 
%The wide format is reserved for long equations
%that cannot easily be set in a single column:
%\begin{widetext}
%\begin{equation}
%{\cal R}^{(\text{d})}=
% g_{\sigma_2}^e
% \left(
%   \frac{[\Gamma^Z(3,21)]_{\sigma_1}}{Q_{12}^2-M_W^2}
%  +\frac{[\Gamma^Z(13,2)]_{\sigma_1}}{Q_{13}^2-M_W^2}
% \right)
% + x_WQ_e
% \left(
%   \frac{[\Gamma^\gamma(3,21)]_{\sigma_1}}{Q_{12}^2-M_W^2}
%  +\frac{[\Gamma^\gamma(13,2)]_{\sigma_1}}{Q_{13}^2-M_W^2}
% \right)\;. 
% \label{eq:wideeq}
%\end{equation}
%\end{widetext}
%This is typed to show how the output appears in wide format.
%(Incidentally, since there is no blank line between the \texttt{equation} environment above 
%and the start of this paragraph, this paragraph is not indented.)

\begin{acknowledgments}
We wish to acknowledge the CHPC under projects MATS862 and MATS0868 for computer time, Kenya Education Network Trust (KENET) research computing services for access to GPU cloud services used in this High Throughput Work. MOA acknowledges discussions with the porE developers.

\end{acknowledgments}

\appendix
\section{Appendixes}
\subsection{\label{app:subsec}Code Availability}
The python code, and databases used in this paper is available online at \url{https://github.com/mikeatm/zif-automation} once the paper is published.

\section{Pore Volume}
\begin{figure*}
\includegraphics[height=60mm,width=1.5\columnwidth]{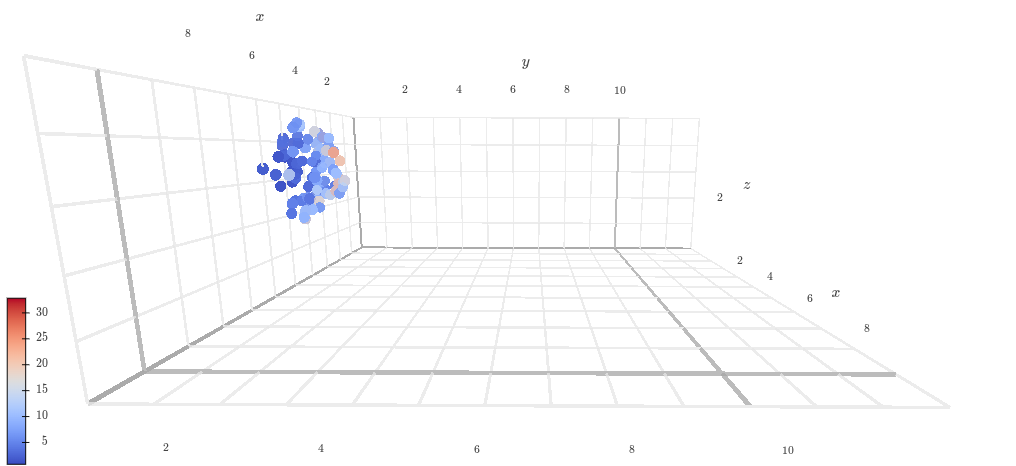}%
\caption{\label{fig:with-context} The adsorption energies showing the pore volume in context within the cell.}
\end{figure*}

The pore volume and center was computed as stated with PoRE\cite{trepte2021pore}, with $2000$ monte carlo steps, the results are summarized in Tab.~\ref{tab:poresize}, Fig.~\ref{fig:with-context} shows the adsorption energies within a cell of equivalent size to the unit cell in ZIF-CO3-1, showing the data in Fig.~\ref{fig:no-context} in context.

\begin{table}
\caption{\label{tab:poresize} PoRE size and coordinates for the largest pore in ZIF-CO3-1, the first row shows results from this work.}
\begin{ruledtabular}
\begin{tabular}{lccc}
 ZIF&  pore size & distribution\% & Coordinates \\ \hline
 ZIF$-$CO3-1 & 2.63\AA &  58.5 & (0.535884    6.156901    3.123292) \\
             & 2.85\AA\cite{zheng2023structural} & &
\end{tabular}
\end{ruledtabular}
\end{table}

\bibliography{apssamp}% Produces the bibliography via BibTeX.

\end{document}